# Testing Kepler’s Hypothesis on the Star of Bethlehem: A Kinematic and Astronomical Analysis of the 7 BCE Jupiter–Saturn Conjunction

Marcel Bodor
*Independent Researcher*
Soyaniquilpan de Juárez, Mexico
*mbjtp2016@gmail.com*

François Bauduin
*Independent Researcher*
Brussels, Belgium

## Abstract

This paper presents an interdisciplinary analysis of the “Star of Bethlehem” narrative described in the Gospel of Matthew (Mt 2:1–12), examining the hypothesis, originally proposed by Johannes Kepler, that the reported phenomenon may be associated with the Jupiter–Saturn conjunction of 7 BCE.

The methodology is based on a systematic comparison between the textual account and independently verifiable astronomical data, including retro-calculated planetary ephemerides, the geometry of the sky as observed from Judea, geographical constraints associated with the Jerusalem–Bethlehem route, and the historical chronology of the reign of Herod the Great. The elements of the narrative are treated as a set of distinct and partially independent constraints required to be satisfied simultaneously within an explicitly falsifiable framework. The analysis is conducted within constrained observational and kinematic conditions, thereby avoiding arbitrary parameter selection.

The analysis indicates that the 7 BCE Jupiter–Saturn conjunction—characterized by its triple occurrence and extended duration—exhibits an apparent motion consistent with key aspects of the reported behavior of the star, including its progression and apparent stopping. In particular, the stationary phase of Jupiter is found to occur within a few days of an independently identified sky–ground kinematic synchronization window, without requiring ad hoc parameter adjustments.

A sensitivity analysis suggests that this compatibility remains stable under reasonable variations of the underlying assumptions. In this context, the Jupiter–Saturn conjunction emerges as a particularly coherent candidate satisfying the set of constraints considered.

This study does not aim to establish a definitive historical identification, but rather to propose a physical and testable framework for evaluating the compatibility of celestial configurations with the narrative. It highlights a convergence between astronomical data and textual constraints, indicating that the account cannot be dismissed as scientifically incompatible solely on the basis of rational analysis.

**Note on conventions.**

Throughout this work, dates are expressed using the BCE/CE convention (Before Common Era / Common Era), equivalent to the traditional BC/AD notation. Unless otherwise specified, all dates are given in the Julian calendar, which was in use during the period considered.

## 1. Introduction

For more than four centuries, a hypothesis originally proposed by Johannes Kepler has remained only partially explored from a modern scientific perspective: the astronomical phenomenon described in the Gospel of Matthew (Mt 2:1–12), commonly referred to as the “Star of Bethlehem,” may be associated with a planetary conjunction, in particular the Jupiter–Saturn conjunction of 7 BCE. First discussed by Kepler in *De Stella nova in pede Serpentarii* (1606), this idea has attracted sustained interest but has rarely been examined within a unified framework simultaneously incorporating astronomical, geographical, chronological, and textual constraints. The present work is specifically concerned with this question: not to introduce an additional hypothesis in an already extensive literature, but to assess whether the Keplerian hypothesis remains consistent when examined using the tools of modern astronomy and kinematic modeling.

The “Star of Bethlehem” has been the subject of numerous interpretations, including comets, novae, supernovae, atmospheric optical phenomena, as well as symbolic or theological readings. Despite this diversity, many existing approaches exhibit at least one of the following limitations: lack of a falsifiable framework, reliance on ad hoc assumptions, or partial analyses that do not simultaneously account for the full set of constraints described in the Matthean narrative. In particular, to our knowledge, no previous study has systematically compared, within a quantitative and kinematic framework, the apparent motion of a celestial phenomenon with a realistic terrestrial displacement such as that described in the account of the Magi.

In the present study, the text of Matthew is treated as a set of minimal descriptive constraints, considered independently of theological interpretation: the appearance of an identifiable celestial object, the perception of its motion (“going before them”), the indication of a geographical direction, and the description of a temporally localized “stopping” above a specific destination. Each of these elements is interpreted as a condition that any explanatory model should satisfy simultaneously. If one or more of these conditions cannot be met, the corresponding hypothesis is considered incompatible within the proposed model.

The aim of this work is not to establish a historical fact in the strict sense, nor to validate a religious belief. Rather, it seeks to evaluate the scientific coherence of an ancient narrative when examined using the tools of modern astronomy, celestial geometry, historical chronology, and probabilistic reasoning. In this context, no hypothesis invoking supernatural intervention is considered, not as a philosophical stance, but because such explanations fall outside the domain of empirical verification.

To clarify the underlying mechanism, the present analysis is based on the following observation: the narrative does not require that the celestial object guide the observers continuously during their movement. Rather, it can be interpreted as describing a configuration in which a celestial body reaches a stationary phase relative to the stellar background, while simultaneously being aligned with the direction of travel at the end of a short terrestrial trajectory.

In this framework, the selected date does not result from an arbitrary choice, but from the intersection of independent constraints — visible after sunset, travel duration, and geometric alignment — which together define a narrow temporal window of compatibility. The proposed mechanism is therefore a kinematic synchronization between celestial configuration and terrestrial motion, rather than a direct guiding effect during the journey.

The central hypothesis examined is that the Jupiter–Saturn conjunction of 7 BCE, characterized by its triple occurrence and unusually long duration, may constitute a candidate capable of satisfying the set of constraints derived from the narrative. This hypothesis is evaluated using astronomical ephemerides, a geographical model of the Jerusalem–Bethlehem route, and a detailed temporal analysis extending to the hourly scale, linking the visibility of the celestial configuration to the kinematics of terrestrial motion.

The originality of this work lies in its integrative and explicitly falsifiable approach. It is worth noting that the initial aim of the study was to test the Keplerian hypothesis critically. The results presented here therefore emerge from a direct comparison between the model and the astronomical data. Rather than relying on cumulative or selectively weighted evidence, the approach is based on a set of restrictive constraints that must be satisfied simultaneously. A probabilistic assessment is also introduced to evaluate the likelihood that such a convergence of constraints could arise by chance within the adopted framework.

Finally, this study is situated within a broader epistemological perspective: it explores the extent to which an ancient textual account may be examined using the methods of modern science without prior reduction to symbolic interpretation or dismissal on non-scientific grounds. In this sense, the question initially raised by Kepler can be framed as a well-defined problem in the history of astronomical interpretation, open to systematic analysis.

### 1.2 Conceptual Overview of the Model

The present approach extends the hypothesis originally proposed by Johannes Kepler by introducing an additional constraint derived from the narrative itself: the terrestrial trajectory from Jerusalem to Bethlehem. A conceptual illustration of this mechanism is provided in Figure A1.

The model is based on a comparison between two coupled dynamics:

(i) the apparent motion of the Jupiter–Saturn conjunction in the sky, and
(ii) the terrestrial motion of a caravan traveling southward from Jerusalem to Bethlehem.

The central question is whether a temporal configuration exists in which these two motions become synchronized, such that the arrival of the travelers coincides with a geometrically significant position of the conjunction — namely, its alignment with the southward direction.

To illustrate this mechanism, one may consider different dates within the period of the triple conjunction. For some dates, the conjunction appears displaced eastward relative to the trajectory at the time of arrival, while for others it has already passed the southern direction. Only within a narrow temporal interval do these constraints converge, yielding a configuration consistent with the narrative description.

In the present model, this condition is satisfied for a date close to late December 7 BCE, for a travel velocity consistent with historical estimates.

Kinematic synchronization between travel and celestial configuration

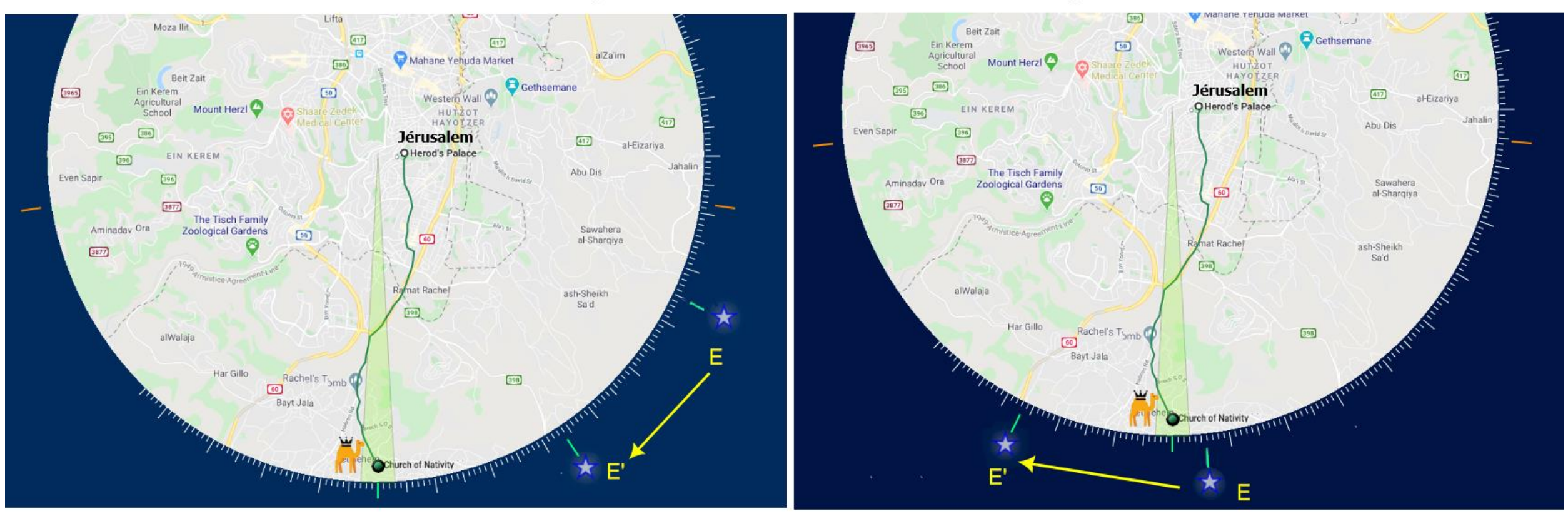

November 15'th 7 BCE
Too early (Jupiter east of alignment at arrival)

January 10'th 7 BCE
Too late (Jupiter west of alignment at arrival)

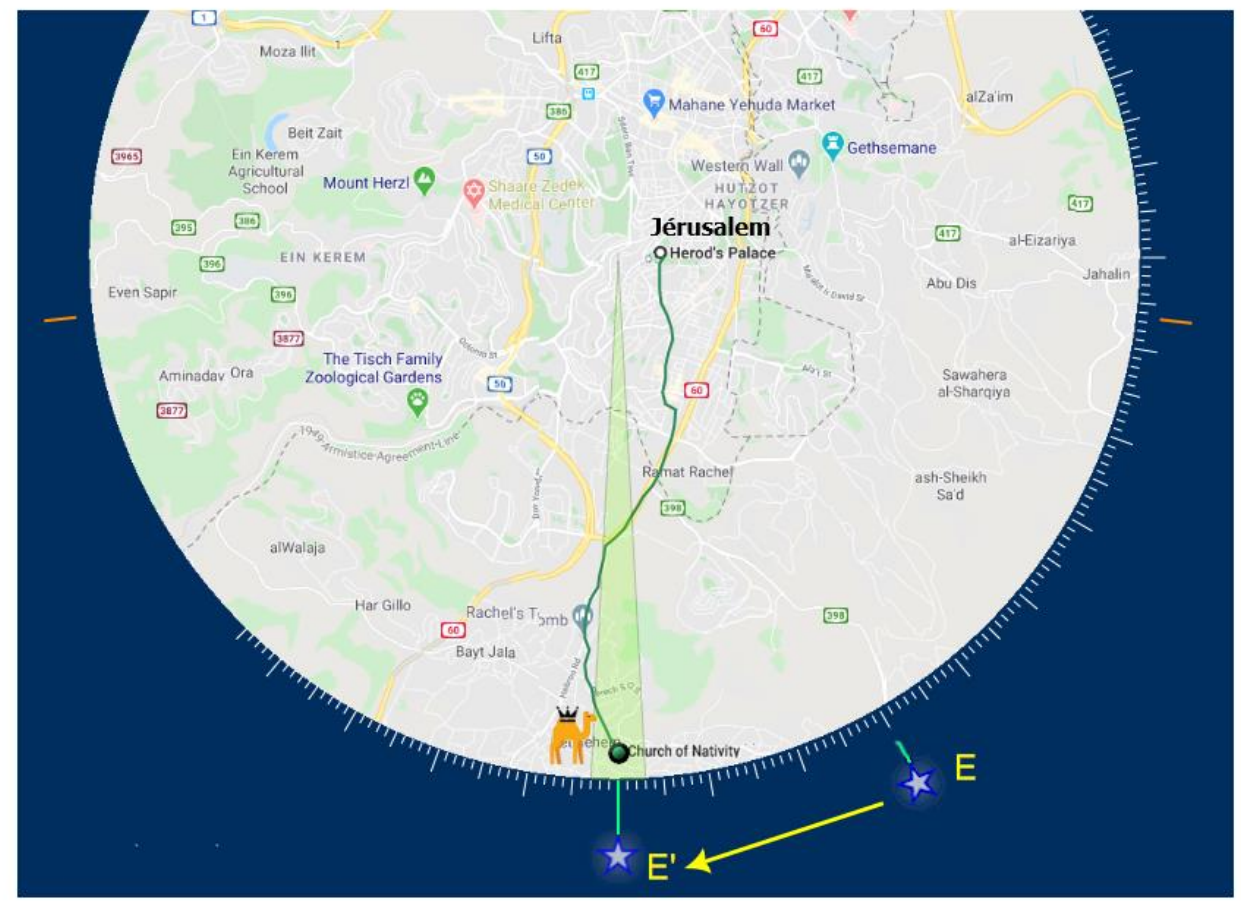

December 24'th 7 BCE
Valid configuration (alignment at arrival)

**Figure A1 — Conceptual illustration of the kinematic synchronization mechanism.**
The three panels represent different dates within the conjunction period. For a given terrestrial travel duration (Jerusalem to Bethlehem), the apparent position of Jupiter at the time of arrival depends on the selected date.
**Left:** if the departure occurs too early, Jupiter remains east of the southward direction at arrival.
**Right:** if the departure occurs too late, Jupiter has already moved westward past the alignment.
**Bottom:** only within a narrow temporal window does the arrival coincide with a configuration in which Jupiter is aligned with the southward direction, satisfying the geometric constraint derived from the narrative.
This illustrates the selection mechanism underlying the results: the valid dates emerge from the intersection of independent constraints rather than from any single astronomical event.

## 2. Historical Context and State of the Question

### 2.1. The Intuition of Johannes Kepler

Johannes Kepler was the first scholar of the early modern period to propose an explicitly astronomical and rational interpretation of the Star of Bethlehem, based on the identification of a real and historically datable planetary configuration. In *De Stella nova in pede Serpentarii* (1606), written following his observation of the supernova of 1604, Kepler suggested that the phenomenon described in the Gospel of Matthew could be associated with a "great conjunction," in particular the conjunction of Jupiter and Saturn in 7 BCE.

Kepler's proposal was not merely symbolic. It relied on a detailed understanding of planetary kinematics and on the best observational data available at the time, notably those derived from the work of Tycho Brahe and compiled in the *Rudolphine Tables*. He noted that the Jupiter–Saturn conjunction of 7 BCE was remarkable in that it was triple and extended over an unusually long period, from the spring of 7 BCE to the beginning of 6 BCE.

However, Kepler's analysis remained incomplete for two main reasons. First, the precision of early seventeenth-century ephemerides did not allow for a detailed modeling of the visibility of the phenomenon from a specific terrestrial location. Second, Kepler lacked a methodological framework for comparing celestial dynamics with a realistic human displacement such as that described in the narrative of the Magi. His hypothesis therefore remained at the level of a plausible conjecture rather than a testable model in the modern scientific sense.

### 2.2. Subsequent Approaches and Their Limitations

Since Kepler, the question of the Star of Bethlehem has been widely discussed, and a broad range of hypotheses has been proposed. These can be grouped into several main categories:

- transient stellar phenomena (novae or supernovae),
- comets,
- planetary conjunctions (notably those proposed by Ferrari d'Occhieppo (1977), who identified the 7 BCE Jupiter–Saturn triple conjunction in Pisces, and Michael R. Molnar (1999), who proposed a lunar occultation of Jupiter in Aries in 6 BCE based on an astrological interpretation),
- atmospheric or optical phenomena,
- symbolic or theological interpretations.

Hypotheses involving novae or supernovae face a significant difficulty: the absence of independent observational records in Chinese, Babylonian, or Greco-Roman sources for the relevant period. Comets, often invoked because of their striking appearance, present another major limitation: their rapid motion and relatively short visibility are difficult to reconcile with a phenomenon described as lasting and interpretable over several months.

Planetary conjunctions have been examined more seriously, particularly in the twentieth century, with contributions from Ferrari d'Occhieppo (1977), Hughes (1979), and Molnar (1999). However, these studies are generally partial. They often focus on temporal coincidence between an astronomical event and a historically plausible period, without systematically evaluating the geometrical and kinematic compatibility with the terrestrial displacement described in Matthew 2:1–12. In particular, the issues of direction ("going before them") and apparent "stopping" are rarely addressed quantitatively.

Finally, many studies conclude that the Matthean narrative is primarily symbolic or theological in nature. While this perspective is legitimate from a hermeneutic standpoint, it tends to bypass the question of factual coherence without fully examining it.

### 2.3. Absence of an Integrated and Falsifiable Model

A common feature of most previous studies is the absence of a framework integrating all relevant constraints simultaneously. Existing analyses typically treat astronomical, geographical, historical, and textual dimensions separately, without testing their mutual compatibility within a unified model.

In particular, the following questions have rarely been addressed in a coordinated manner:

– Could a given astronomical phenomenon have been perceived as exceptional by knowledgeable observers?
– Was its duration compatible with the preparation and execution of a long-distance journey?
– Did its apparent motion allow for a directional correspondence with a specific terrestrial route?
– Was there a restricted temporal window satisfying all these conditions simultaneously?

The absence of coordinated answers to these questions largely explains why Kepler's hypothesis has remained neither conclusively validated nor refuted.

### 2.4. Positioning of the Present Work

The present study explicitly builds upon the Keplerian hypothesis while distinguishing itself through its methodological framework. Its aim is neither to propose an additional hypothesis nor to offer a symbolic reinterpretation of the biblical text, but to subject a well-defined hypothesis to a set of explicit and constraining tests based on verifiable data.

By combining modern astronomical ephemerides, precise geographical modeling, and an explicit probabilistic analysis, this work seeks to address a persistent methodological gap in the literature. It thus offers a systematic evaluation of a question first formulated more than four centuries ago, but not previously examined within a fully integrated framework.

As noted by Barthel and van Kooten, the triple occurrence of the Jupiter–Saturn conjunction in 7 BCE represents a more distinctive astronomical event than an isolated conjunction. However, these authors also identify a major limitation of such interpretations, namely the absence of geographical and astrological elements allowing the precise localization of the event described in the narrative: "*the main drawback of this interpretation remains that this theory lacks geographical-astrological elements that explain where the predicted event is taking place*" (Barthel & van Kooten, 2015, p. 600).

The present work explicitly addresses this limitation by proposing a joint modeling of celestial kinematics and the terrestrial Jerusalem–Bethlehem route, thereby providing a spatial and directional framework consistent with the narrative.

It is worth noting that this limitation has not led to the abandonment of the Keplerian hypothesis. For instance, Owen Gingerich presented a detailed account of Kepler's theory, while David W. Hughes, in his comprehensive review of the various proposed explanations, expressed continued support for the Jupiter–Saturn triple conjunction interpretation:

"David Hughes went into substantial detail in his review of the various theories and declared his ongoing support for the Jupiter–Saturn triple conjunction theory" (Barthel & van Kooten, 2015, p. 647).

## 3. Methodology and Analytical Framework

### 3.1. General Methodological Principles

The aim of this work is not to interpret an ancient text in light of a preconceived hypothesis, but to subject a specific hypothesis to a set of independent and verifiable constraints. The methodology is based on a simple principle: each descriptive element of the Matthean narrative (Mt 2:1–12) is treated as a factual constraint that the model should satisfy simultaneously. Failure to satisfy any single constraint is taken as evidence that the corresponding hypothesis is not supported under this approach.

This methodological approach follows an "all-or-nothing" logic, analogous to that used in many areas of the exact sciences when multiple necessary conditions must be fulfilled for a model to be considered consistent. It is therefore not a cumulative argument in a weak probabilistic sense, but an approach based on global coherence.

### 3.2. Working Hypotheses

Two minimal working hypotheses are provisionally adopted:

1. The narrative in Matthew 2:1–12 describes events presented as real within the text, independently of later theological interpretations.
2. The laws governing planetary motion, as formulated by Johannes Kepler and refined by modern celestial mechanics, were already valid at the time considered and can be applied retrospectively to compute planetary positions.

These hypotheses are not treated as ideological assumptions. They are maintained only insofar as no internal contradiction or incompatibility with verifiable data arises.

Taken together, they constitute the sole interpretative framework adopted throughout the analysis.

Dates are expressed in the Julian calendar used during the period considered. Unless otherwise specified, all reported times correspond to local apparent solar time reconstructed from the astronomical software configurations described in Section 9.

### 3.3. Scope of the Study

The scope of the analysis is deliberately restricted in order to avoid methodological ambiguity.

From a textual standpoint, the analysis is limited strictly to the passage of the Gospel of Matthew concerning the Magi (Mt 2:1–12). No elements from other biblical or apocryphal texts are introduced to support the argument.

From an explanatory standpoint, any hypothesis invoking direct supernatural intervention (miracles, angelic appearances, or ad hoc celestial phenomena) is explicitly excluded. Such interpretations, by their nature, cannot be incorporated into a falsifiable scientific framework.

The phenomenon under investigation is therefore assumed to be entirely natural, observable, and governed by known physical laws.

### 3.4. Constraints Derived from the Matthean Narrative

The text of Matthew[1] imposes a set of explicit or implicitly inferred constraints that structure the analysis:

– the existence of an identifiable and observable celestial phenomenon;
– a level of significance sufficient to justify a deliberate long-distance journey;
– a duration compatible with repeated observation, interpretation, and travel;
– the possibility of directional guidance ("going before them");
– a final spatial correspondence between the apparent position of the object and a specific

[1] The constraints are derived directly from the critical Greek text (Nestle–Aland, 28th edition), independently of any confessional translation. The passage Mt 2:1–12 is cited from: Novum Testamentum Graece, Nestle–Aland, 28th ed., Deutsche Bibelgesellschaft, Stuttgart.

terrestrial location ("stood over the place");
– chronological compatibility with the reign of Herod the Great.

These constraints are not derived from a symbolic reading of the text, but from a minimal formalization of its descriptive content in geometrical, kinematic, and chronological terms.

#### 3.4.1. Independence and Informational Orthogonality of Constraints

The constraints extracted from the text are not treated as a homogeneous narrative block, but as a set of distinct observational statements.

Each constraint corresponds to a different aspect of the described phenomenon: temporal evolution, apparent direction, and kinematic behavior (including apparent stopping). As such, they can be considered partially independent in informational terms.

While strict statistical independence cannot be formally demonstrated due to the textual nature of the data, the diversity of these constraints limits the risk of redundancy or circular reasoning. Their simultaneous satisfaction therefore requires a coherent underlying phenomenon rather than arbitrary adjustment.

In this sense, their combined use constitutes a strong constraint on any explanatory model, significantly reducing the space of compatible solutions.

### 3.5. Celestial Modeling

The celestial model is based on modern ephemerides of the Jupiter–Saturn conjunction of 7 BCE. The parameters considered include:

– apparent angular positions of the planets;
– their relative motion over time;
– times of rising, meridian transit, and setting as observed from the region of Judea;
– the duration and structure of the triple conjunction.

These data allow a detailed reconstruction of the apparent kinematics of the phenomenon as perceived by terrestrial observers.

### 3.6. Terrestrial Modeling

The terrestrial model concerns the displacement of the Magi between Jerusalem and Bethlehem. It is based on:

– the geographical distance between the two locations;
– a range of velocities consistent with caravan travel over moderately uneven terrain;
– the directional axis of the route.

The model assumes a departure from Jerusalem at a time corresponding to the first visibility of the phenomenon (early-evening visibility)[2], followed by an arrival in Bethlehem after a time interval determined by the assumed velocity.

### 3.7. Central Condition of Correspondence

[2] Early-evening visibility refers to the first detectable appearance of a celestial object after sunset, when it emerges from solar glare under local twilight conditions. This term is used here in a purely observational sense and is not related to the classical definition of heliacal rising, which applies to pre-sunrise conditions for stars.

The core of the analysis is a synchronization condition between terrestrial motion and celestial motion: the time required for the caravan to travel from Jerusalem to Bethlehem should be comparable to the time taken by the celestial configuration to evolve from its first visibility to a position aligned with the southern meridian, corresponding to the direction of Bethlehem as seen from Jerusalem.

This condition is tested systematically over the full visibility period of the conjunction. The only free parameter in the model is the travel velocity, constrained to remain within realistic bounds.

This parameter is explored within a plausible range and is not finely tuned to produce a specific outcome.

### 3.8. Robustness to Auxiliary Assumptions

The model relies on several auxiliary assumptions, including travel velocity, route selection, and observational conditions.

These are not finely adjusted parameters, but rather plausible values consistent with historical and physical knowledge. A sensitivity analysis indicates that reasonable variations in these assumptions do not significantly affect the temporal coherence or the general conclusions of the model.

In particular, the geometrical and kinematic constraints derived from the celestial configuration remain dominant, such that auxiliary assumptions do not introduce critical dependencies.

The model should therefore be understood as structurally robust rather than dependent on a specific parameter choice.

### 3.9. Falsifiability and Reproducibility

The approach is explicitly falsifiable. The model may be considered unsupported if:

– no date satisfies all constraints simultaneously;
– required parameters fall outside realistic ranges;
– results depend on arbitrary adjustments.

To ensure reproducibility, a computational tool was developed to allow independent testing of the model conditions using the same ephemerides and geographical parameters. The results therefore rely on transparent calculations rather than on authority.

### 3.10. Nature and Scope of the Result

This work does not aim to establish a definitive identification of the Star of Bethlehem.

Rather, it proposes a physical and testable framework for evaluating the compatibility of celestial configurations with the constraints derived from the narrative.

Within this framework, the Jupiter–Saturn conjunction emerges as a particularly robust candidate, in that it satisfies the set of constraints considered without requiring arbitrary adjustments.

It significantly reduces the space of compatible solutions, while not implying strict uniqueness.

The result should therefore be interpreted as a convergence of independent constraints rather than as a formal demonstration of uniqueness, and as an argument for the physical plausibility of this configuration without excluding the possibility of alternative compatible solutions.

## 4. Astronomical Analysis of the Jupiter–Saturn Conjunction of 7 BCE

### 4.1. Definition and General Framework

A planetary conjunction refers to the apparent configuration in which two planets exhibit a small angular separation as seen from Earth. Conjunctions between Jupiter and Saturn—so-called "great conjunctions"—occur with an average periodicity of approximately twenty years. However, certain configurations stand out due to their kinematic complexity and duration, particularly when they take the form of a triple conjunction associated with the apparent retrograde motion of the planets.

The Jupiter–Saturn conjunction of 7 BCE belongs to this specific category and may be regarded as one of the more remarkable planetary configurations observable in antiquity.

### 4.2. Structure of the Triple Conjunction

The Jupiter–Saturn conjunction of 7 BCE does not correspond to a single instantaneous event but to an extended astronomical configuration.

Figure 1 illustrates the evolution of the apparent angular separation between Jupiter and Saturn over a period of approximately 300 days, from the spring of 7 BCE to the beginning of 6 BCE.

The curve exhibits three successive minima in angular separation, corresponding to the three characteristic approaches of a triple conjunction. This structure arises from the apparent retrograde motion of Jupiter, resulting from the combined orbital motions of Earth, Jupiter, and Saturn.

Three distinct conjunctions can thus be identified:

– a first conjunction in late May;
– a second conjunction in late September;
– a third conjunction in early December.

This sequence gives the 7 BCE conjunction its distinctive character—not due to an exceptionally small angular separation, but due to its extended duration and kinematic complexity.

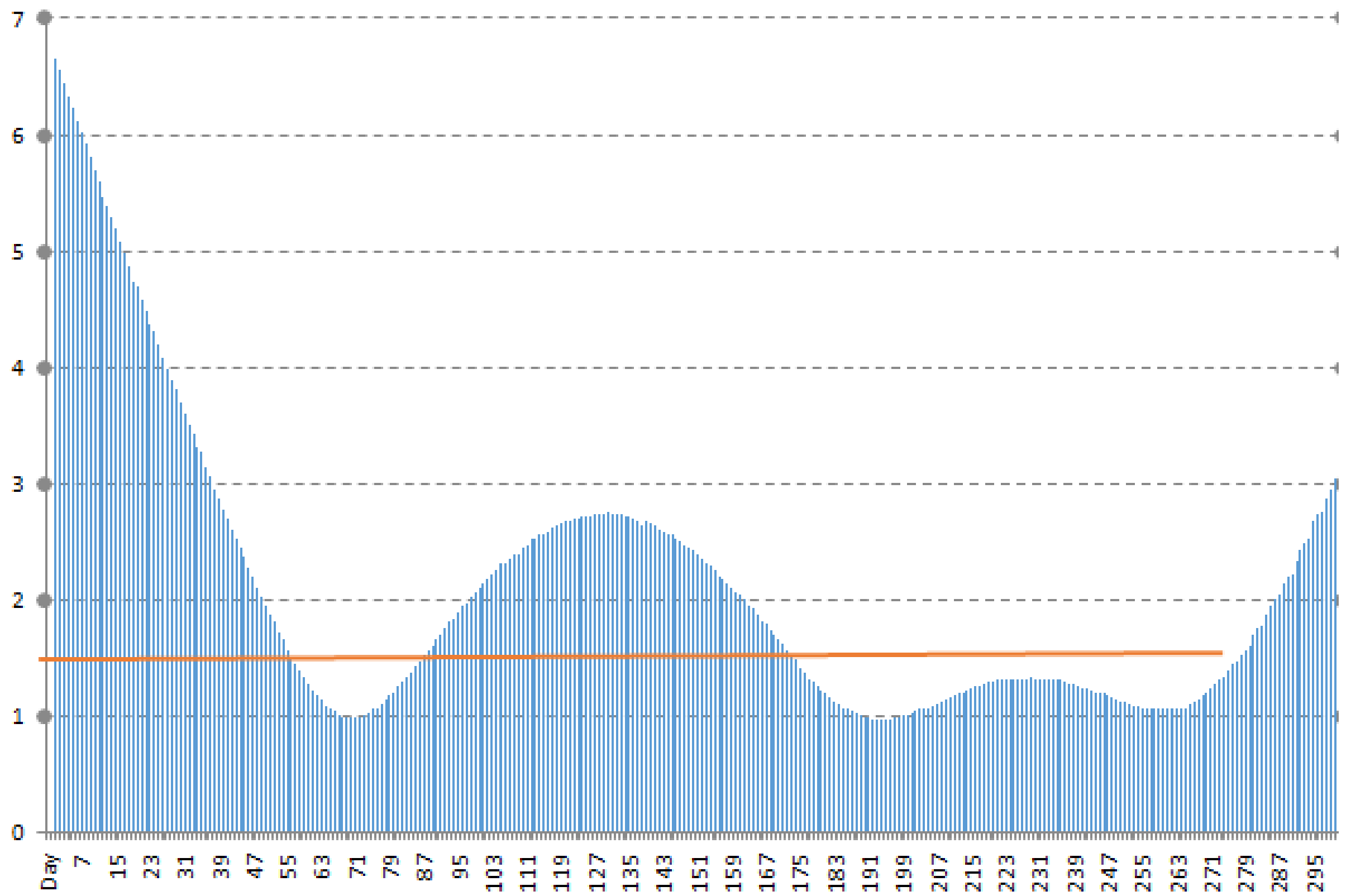

**Figure 1 — Evolution of the apparent angular separation between Jupiter and Saturn during the great conjunction of 7 BCE.**
The figure shows the angular separation between Jupiter and Saturn, expressed in degrees, as a function of time, measured in days from 21 March 7 BCE (Julian calendar). The three successive minima correspond to the three close approaches characteristic of the triple conjunction, resulting from the apparent retrograde motion of Jupiter. This exceptional configuration extends over a period of approximately 300 days, from March 7 BCE to January 6 BCE (Meeus, Astronomical Algorithms, 1998).

### 4.3. Exceptional Duration of the Phenomenon

Unlike most visually striking celestial phenomena—such as comets, meteors, or eclipses—the Jupiter–Saturn conjunction of 7 BCE is not a brief or transient event. Instead, it extends over a period of roughly 300 days, during which the two planets remain sufficiently close in angular separation to be perceived as associated.

This extended duration distinguishes it from most alternative hypotheses and may be considered compatible with the temporal constraints implied by the Matthean narrative (Mt 2:1–12), which suggests:

- an initial observation;
- a phase of interpretation;
- preparation for a journey;
- and a subsequent observation after travel.

Within this context, planetary stations[3] may have contributed to the perception of an evolving and persistent phenomenon. While a single station is not uncommon, the presence of multiple stations within a prolonged configuration may have reinforced the impression of an unusual and sustained celestial event.

## 4.4. Visibility from the Region of Judea

The Jupiter–Saturn conjunction of 7 BCE would have been observable from the latitudes of Judea under generally favorable conditions.

During this period, the planets are located in the zodiacal region of Pisces, at a declination compatible with sustained visibility from the Near East. They rise in the east to southeast, culminate in the southern sky, and set in the west.

Their apparent brightness, significantly exceeding that of background stars, makes them readily visible to the naked eye under typical observing conditions. No specialized instrumentation would have been required.

This accessibility supports the plausibility of repeated observation over extended periods and is consistent with the possibility of interpretation by observers familiar with planetary motion.

## 4.5. Apparent Kinematics and Interpretation of Motion

A frequently discussed element of the Matthean narrative is the statement that the star "went before them" (Mt 2:9). From an astronomical perspective, this description may be interpreted in terms of the apparent motion of a celestial object as observed from Earth, without requiring a non-natural explanation.

In 7 BCE, Jupiter exhibits apparent retrograde motion, a well-known consequence of the relative geometry of Earth's and Jupiter's orbits. During this phase, the planet's apparent angular velocity decreases, approaches zero near stationary points, temporarily reverses direction, and then resumes direct motion.

To an observer, this behavior corresponds to a slow, continuous, and directional motion, interspersed with phases during which the planet appears nearly stationary.

Such kinematic behavior differs from that of most transient celestial phenomena and may be described, in non-technical language, as an object "progressing" across the sky.

In the present analysis, the term "apparent motion" refers to Jupiter's angular motion relative to the stellar background (sidereal frame), not to its diurnal motion with respect to the local horizon induced by Earth's rotation. This distinction is essential for interpreting the stationary phase discussed below. Jupiter becomes observable shortly after sunset during the relevant period, which restricts the effective observation window to the early evening.

In the present framework, the relevant observational condition is approximated by the onset of civil twilight, corresponding to the first practical visibility of Jupiter after sunset. This choice constitutes a physically motivated but not unique interpretation of the narrative, and the results should be understood within this assumption.

[3] A planetary station corresponds to the apparent moment at which a planet's motion reverses direction (from direct to retrograde motion, or vice versa), resulting in a near-zero apparent angular velocity as seen from Earth.

In the present implementation, first practical visibility is approximated by the moment at which Jupiter becomes visually detectable in the simulated sky under civil twilight conditions. This criterion relies on the calibrated visual rendering implemented in Starry Night Pro and is used here as an operational observational threshold rather than as a detailed atmospheric extinction model.

## 4.6. Planetary Station and Apparent "Stopping"

During planetary stations—periods when the apparent angular velocity of a planet approaches zero—the object appears nearly motionless in the sky for several consecutive days.

This well-documented astronomical phenomenon provides a possible natural interpretation of the description that the star "stopped," in the sense of an apparent cessation of motion.

The stationary phase corresponds to the epoch at which the time derivative of Jupiter's apparent longitude approaches zero, i.e., when its angular velocity relative to the stellar background becomes minimal in the sidereal frame.

In the case of the 7 BCE conjunction, this stationary phase occurs in temporal proximity to the third conjunction. According to the ephemerides used in this study, the apparent reversal of Jupiter's motion—defining the stationary phase—occurs around 20–21 December 7 BCE.

This timing lies within a few days of the independently identified kinematic synchronization window discussed in Section 7.7. The proximity of these two independently derived results may be regarded as an element of internal consistency within the model.

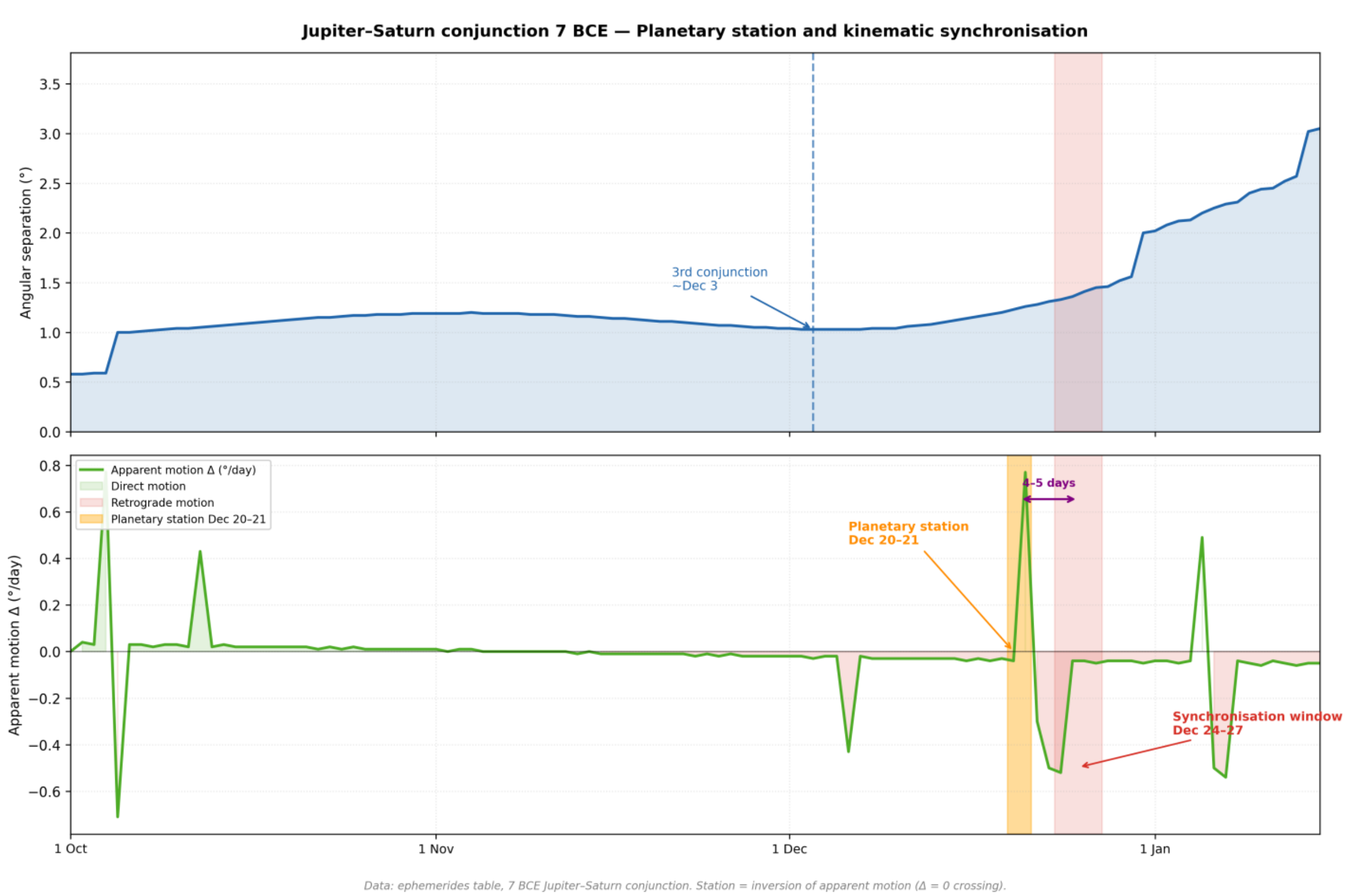

**Figure 2 — Planetary station of Jupiter and kinematic sky–earth synchronization window (October 7 BCE – January 6 BCE).**

**Top panel:** evolution of the apparent angular separation between Jupiter and Saturn, expressed in degrees. The minimum observed around 1–6 December corresponds to the third conjunction of the triple conjunction of 7 BCE.
**Bottom panel:** daily variation (Δ) of the apparent azimuthal position of Jupiter, expressed in degrees per day. Zero crossings of Δ indicate transitions between direct and retrograde motion (stationary points). The stationary phase corresponds to the interval during which Δ remains close to zero, reflecting minimal angular velocity in the sidereal frame; this phase is observed around 20–21 December (orange region).

The red region indicates the kinematic sky–earth synchronization window (24–27 December), identified independently in Section 7.7. The 4–5 day interval separating these two events, obtained from independent analyses based on the same ephemerides, constitutes an element of internal consistency of the model. When read in conjunction with Figure 1, this figure highlights what makes the late-December period uniquely exceptional within the overall ~300-day conjunction.

## 4.7. Comparison with Alternative Astronomical Hypotheses

Various hypotheses have been proposed to explain the phenomenon described in Matthew, including comets, novae, supernovae, and alternative planetary configurations.

However, many of these proposals do not simultaneously satisfy the full set of constraints suggested by the narrative, including:

– extended visibility;
– directional coherence during motion;
– the presence of an apparent stationary phase;
– chronological compatibility with the reign of Herod the Great;
– repeated naked-eye visibility.

In this context, the Jupiter–Saturn conjunction of 7 BCE may be considered a particularly consistent candidate with respect to these combined criteria.

## 4.8. Exclusion of Adjacent Jupiter–Saturn Conjunctions Based on Duration

The hypothesis explored in this work focuses on the 7 BCE conjunction as a candidate satisfying the set of constraints derived from the narrative. It is therefore relevant to examine why adjacent Jupiter–Saturn conjunctions—such as those of 26 BCE and 15 CE—do not provide comparable candidates.

A typical (single) conjunction produces only one close approach between the two planets and lasts on the order of a few weeks. It does not involve prolonged retrograde motion, multiple stationary phases, or extended visibility over several months.

By contrast, the narrative implies a sequence of events extending over a longer timescale, including observation, interpretation, travel, and subsequent observation.

A short-lived conjunction would not remain observable throughout such a sequence. In this sense, only a triple conjunction—resulting from extended retrograde motion and lasting several months—may be considered compatible with this temporal constraint.

The 7 BCE conjunction satisfies this condition, whereas adjacent simple conjunctions do not. This criterion is derived directly from the narrative constraints and does not depend on geometric or kinematic fine-tuning.

It also contributes to the rarity factor discussed in Section 8, as triple Jupiter–Saturn conjunctions occur on timescales of the order of eight to nine centuries and represent a distinct class of events compared to ordinary conjunctions.

## 5. Celestial Geometry and Directional Correspondence between Jerusalem and Bethlehem

### 5.1. The Problem of Directional Guidance

One of the most debated elements of the Matthean narrative concerns the statement that the observed object “went before” the Magi and guided them to Bethlehem. This formulation has often led to symbolic or supernatural interpretations, due to the apparent difficulty of identifying an astronomical phenomenon capable of indicating a precise geographical direction over a short terrestrial distance.

Any astronomical interpretation of the phenomenon described in Matthew must therefore address a fundamental geometrical question: can an observable celestial object provide a usable directional indication during real human movement, without invoking a miraculous or purely metaphorical explanation?

Building on the kinematic properties discussed in Section 4.5, the analysis shifts here from the astronomical to the geometrical and terrestrial domain. The objective is no longer to describe the motion of the celestial object itself, but to evaluate how such motion could be perceived and interpreted by an observer moving along a constrained terrestrial path.

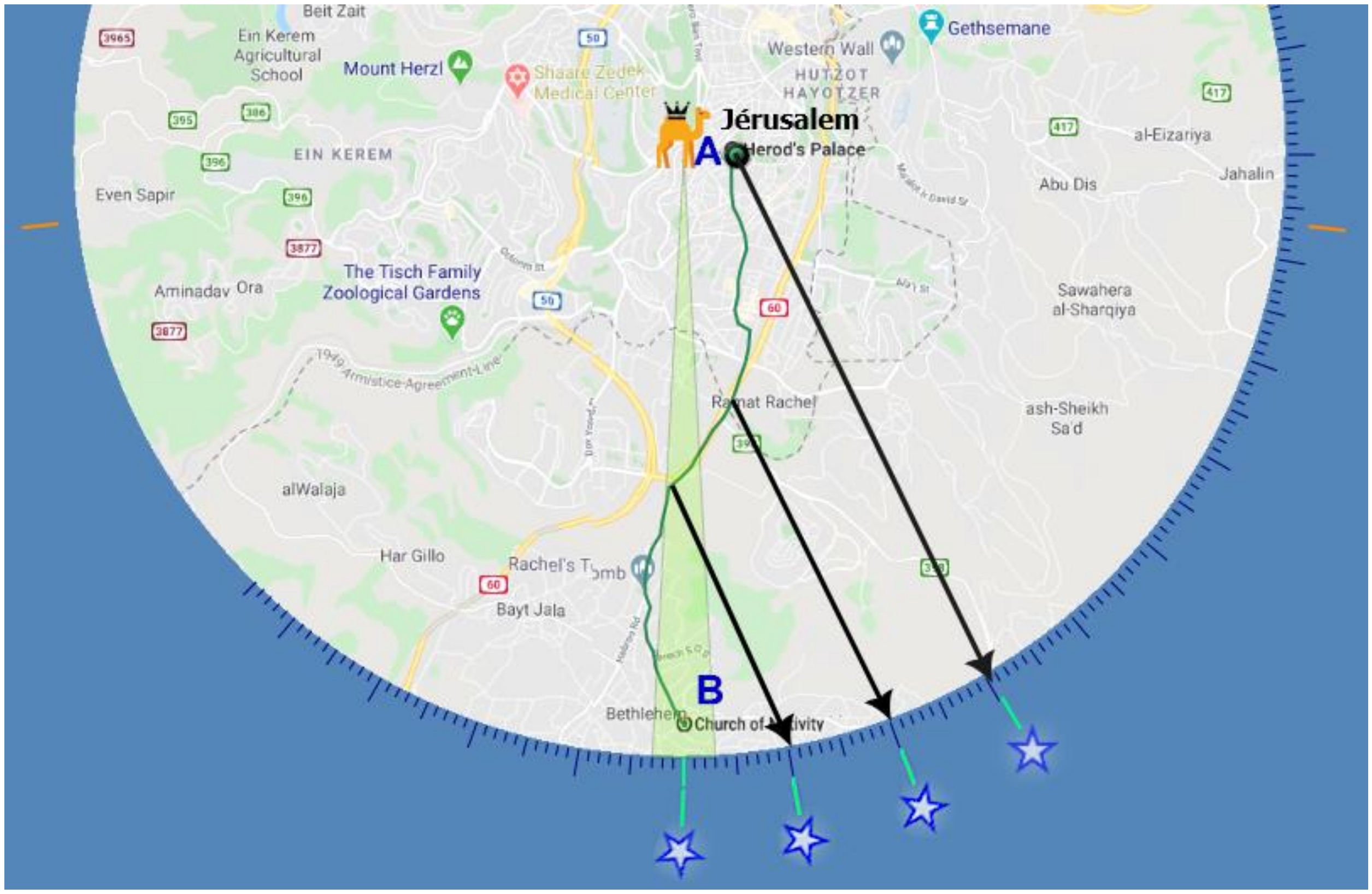

**Figure 3 — Schematic correspondence between the apparent motion of Jupiter along the southern horizon and the terrestrial displacement along the Jerusalem–Bethlehem route.**

The apparent trajectory of the celestial object, as observed at human eye level, maintains a stable direction relative to the axis of travel, allowing for a directional interpretation without invoking any supernatural assumption.

## 5.2. Geographical and Kinematic Constraints of the Jerusalem–Bethlehem Route

The route connecting Jerusalem to Bethlehem is oriented approximately along a north–south axis, with a slight eastward component. The distance between the two locations is about 8.5 km. Assuming an average travel speed on the order of 5–6 km·h$^{-1}$—consistent with historical and ethnographic data for caravan travel—the journey duration is on the order of 1 to 2 hours[4].

This estimate is intended as an order-of-magnitude approximation; a more refined range is discussed in Section 7.3.

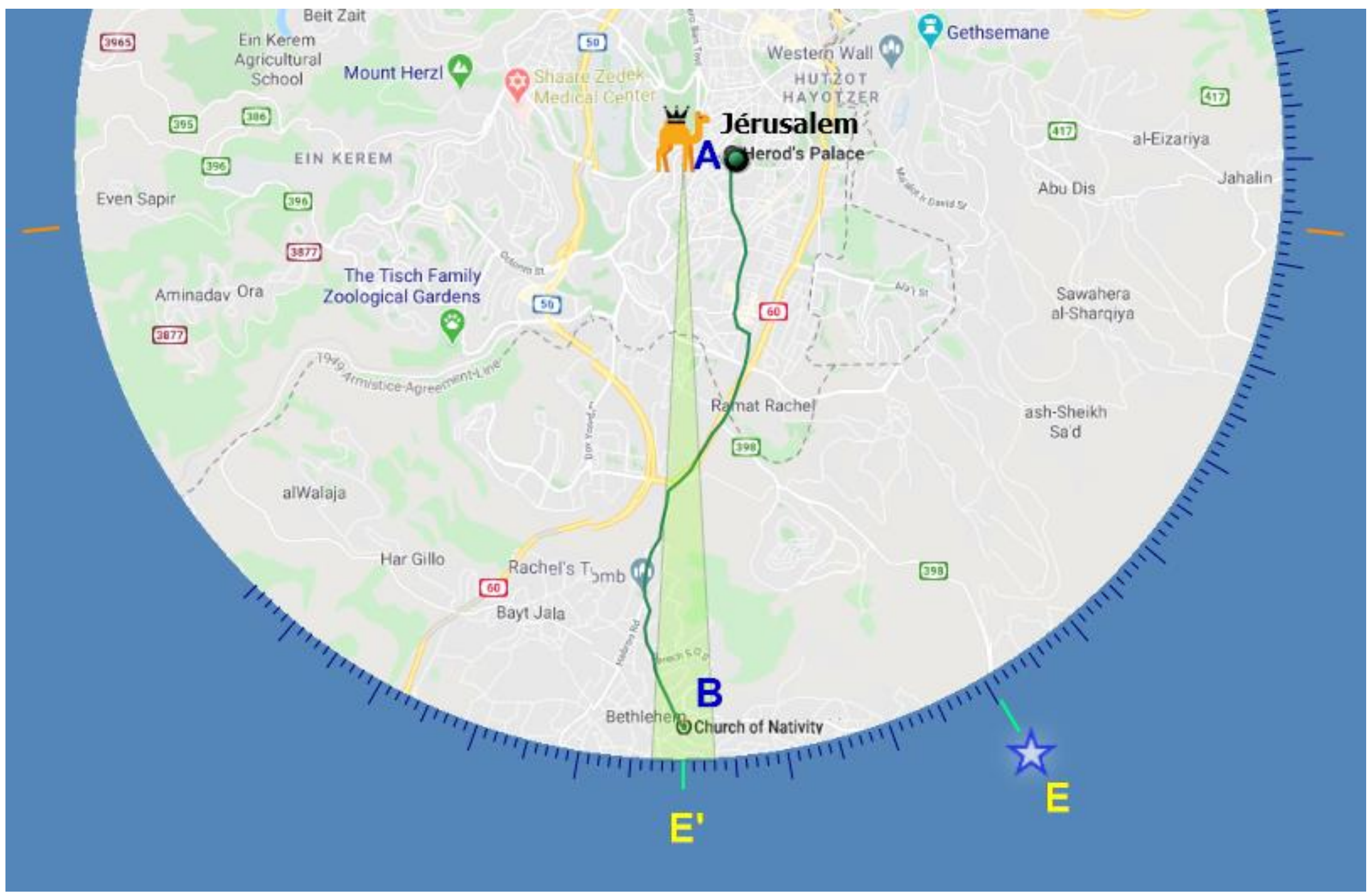

**Figure 4 — Caravan route from Jerusalem to Bethlehem along the ancient Hebron road.**
A–B displacement, predominantly oriented along a north–south axis, corresponds to a displacement of approximately 8.5 km. This orientation implies, for a moving observer, a direction of terrestrial progression aligned with the southern sector of the sky. The figure also illustrates the apparent position of the Jupiter–Saturn conjunction at the time of its early-evening visibility, highlighting the geometric consistency between the direction of travel and the location of the celestial phenomenon.

For an observer departing from Jerusalem toward Bethlehem, the direction of motion corresponds approximately to the southern sector of the sky. Any astronomical candidate must therefore be compatible with the presence of a visible object in that region, whose apparent position could be perceived as lying "ahead" of the direction of travel.

[4] Typical caravan travel speeds in antiquity are generally estimated between 4 and 7 km·h$^{-1}$, depending on terrain, load, and travel conditions. This range is consistent with historical and archaeological evidence, including data from Near Eastern trade routes (e.g., Kültepe tablets) and zootechnical studies on camel locomotion. The value of 6 km·h$^{-1}$ adopted in this work corresponds to the upper bound of this range and is appropriate for short-distance travel under moderate conditions. The robustness of the model has been verified across the full interval 4–7 km·h$^{-1}$.

This imposes a simple but restrictive condition: the celestial phenomenon must not only be visible in the southern sky, but must also exhibit an apparent evolution compatible with the duration of the terrestrial journey. Any hypothesis leading to a significant mismatch between travel time and celestial evolution would not be supported in this context.

### 5.3. Apparent Position of the Conjunction in the Sky

From the time of its first early-evening visibility, the Jupiter–Saturn conjunction remains located in the southern sector of the sky for an observer in Judea.

After rising in the east to southeast, the planets culminate in the southern sky before setting in the west, in accordance with the apparent motion of zodiacal objects at these latitudes.

This configuration has an important geometrical implication: during a southward terrestrial displacement in the evening or early night, the apparent position of the conjunction remains broadly aligned with the direction of motion, without significant lateral deviation.

In this sense, the celestial configuration may be perceived as remaining "in front" of the observer, in a qualitative directional sense, without requiring any non-natural interpretation.

### 5.4. Effect of Apparent Motion on Directional Perception

The slow apparent motion of the planets involved in the conjunction, combined with the extended duration of the phenomenon, allows repeated observations without abrupt positional changes.

Unlike fast-moving or transient celestial objects, the conjunction maintains sufficient spatial coherence to be associated, over days or weeks, with a relatively stable reference in the sky.

For observers familiar with planetary cycles, such as those in antiquity, the apparent progression of the conjunction—interpreted within an astrological or symbolic framework—may have been perceived as carrying directional significance.

This interpretation does not require any additional assumptions beyond the known kinematic properties of the phenomenon and historically attested observational practices.

### 5.5. The "Stopping" Phenomenon and Final Localization

The Matthean narrative states that the object "stopped over the place where the child was." This introduces a strong descriptive constraint requiring a compatible physical interpretation.

From an astronomical perspective, an apparent "stopping" naturally corresponds to a planetary station, during which the apparent angular velocity of a planet becomes very small before reversing direction. During this phase, the position of the object remains nearly unchanged over successive nights.

For an observer arriving at a destination, the coincidence between the end of the terrestrial displacement and a phase of minimal apparent motion—particularly near meridian transit—may be perceived as the object "standing over" the location reached.

This interpretation does not imply strict geometrical verticality or precise pointing in the modern sense. Rather, it reflects a phenomenological description, expressed in ordinary language, based on temporal and directional coincidence at the scale of human perception.

### 5.6. Temporal Compatibility with Arrival in Bethlehem

The stationary phase associated with the third Jupiter–Saturn conjunction occurs within a time window that may be considered compatible with a displacement from Jerusalem to Bethlehem following the meeting with Herod described in the narrative.

Within the proposed framework, the same celestial phenomenon may account—without discontinuity—for the initial appearance, the directional role during travel, the apparent stationary behavior at arrival.

This continuity removes the need to postulate multiple distinct phenomena or introduce ad hoc adjustments to explain different stages of the narrative. The temporal coherence of the configuration may therefore be regarded as a structural feature of the model.

### 5.7. Limits and Precision of the Directional Interpretation

It is important to emphasize that the proposed model does not assume precise localization in the modern sense, based on exact geographical coordinates or strict vertical projection.

The guidance considered here is primarily directional and qualitative, consistent with known practices of orientation and travel in antiquity.

Within the proposed model, the correspondence between the celestial geometry of the Jupiter–Saturn conjunction and the geography of the Jerusalem–Bethlehem route appears sufficient to satisfy the descriptive constraints of the Matthean text, without exceeding what an ancient observer—attentive but unaided by instruments—could reasonably perceive and interpret.

The model therefore does not imply strict or unambiguous directional guidance, but rather compatibility with a plausible directional interpretation in an ancient observational context.

## 6. Temporal Window and Chronological Constraints

### 6.1. The Problem of Dating

Any hypothesis seeking to identify the Star of Bethlehem with a real astronomical phenomenon must satisfy a set of independent chronological constraints, derived both from the Gospel of Matthew and from historical data.

The issue is not only whether a remarkable celestial event occurred, but whether it falls within a time frame compatible with the sequence of events described in the narrative.

The aim of this section is therefore to examine the temporal consistency between the Jupiter–Saturn conjunction of 7 BCE and the historical context of the reign of Herod the Great, without presupposing the validity of either source. The objective is to determine whether a non-contradictory chronological window can be identified.

### 6.2. Textual Constraints from Matthew

The Gospel of Matthew explicitly places the birth of Jesus "in the days of King Herod." This provides a firm upper chronological bound, as Herod the Great is generally understood to have died in 4 BCE, based on historical sources such as Flavius Josephus.

The narrative also implies a non-negligible interval between the initial observation of the phenomenon by the Magi, their journey, their arrival in Jerusalem, and their subsequent travel to Bethlehem.

Such a sequence is consistent with a celestial phenomenon extended in time and observable on multiple occasions, but is difficult to reconcile with short-lived events such as rapidly moving comets or transient novae.

## 6.3. Chronology of the 7 BCE Conjunction

Astronomical ephemerides indicate that the Jupiter–Saturn conjunction of 7 BCE exhibits a triple-conjunction structure, resulting from the apparent retrograde motion of Jupiter relative to the background stars.

This mechanism produces three successive close approaches between the two planets, occurring in spring, autumn, and early winter of the same year.

As a result, the phenomenon extends over a period of approximately 300 days, during which Jupiter and Saturn remain in sufficient angular proximity to be perceived as associated.

This temporal persistence distinguishes the event from most other proposed candidates and may be considered compatible with prolonged observation and sustained interpretation within ancient observational frameworks.

## 6.4. Synchronization with the Narrative Sequence

The sequence described in Matthew—initial observation, travel toward Jerusalem, meeting with Herod, and subsequent journey to Bethlehem—implies a temporal structure that may be compared with the observable phases of the proposed astronomical phenomenon.

Under this approach, the third conjunction, occurring toward the end of 7 BCE and associated with a planetary stationary phase, corresponds to a configuration in which the celestial object may be perceived both as accompanying a southward movement and as exhibiting an apparent stopping near the end of the journey.

The temporal proximity between this phase of the conjunction and the sequence described in the text may therefore be regarded as contributing to the overall internal consistency of the proposed interpretation.

## 6.5. Exclusion of Alternative Periods

The first two conjunctions of 7 BCE, although astronomically significant, present several limitations within the present framework:

– they occur at times less compatible with a southward evening displacement from Jerusalem;
– they are not associated with a planetary stationary phase that could correspond to an apparent cessation of motion;
– they do not provide a unified account of the full set of narrative expressions.

In addition, examination of adjacent years (8 and 6 BCE) does not reveal planetary configurations combining extended duration, repeated visibility, and geometrical compatibility with the directional and temporal constraints discussed above.

### 6.6. Robustness of the Temporal Window

The temporal window identified around the end of 7 BCE does not arise from arbitrary adjustment, but from the convergence of multiple independent constraints:

– well-established astronomical data;
– externally documented historical chronology;
– explicit narrative requirements derived from the text of Matthew.

This convergence significantly reduces the space of compatible solutions and provides a degree of temporal robustness to the proposed framework.

### 6.7. Interpretation of the Chronological Result

It is important to emphasize that the present analysis does not aim to establish with certainty the historical date of the birth of Jesus, but rather to identify a plausible temporal window for the visit of the Magi (Mt 2:1–12).

The objective is more limited, yet methodologically central: to show that the Jupiter–Saturn conjunction of 7 BCE defines a time frame that is consistent with the available astronomical, historical, and textual constraints, without internal contradiction.

A more precise dating would require a finer modeling of observational conditions and of the Jerusalem–Bethlehem displacement, incorporating the apparent kinematics of the celestial configuration during the stationary phase. This analysis is developed in the following section.

## 7. Kinematic Modeling of the Jerusalem–Bethlehem Journey

### 7.1. Problem formulation

The core hypothesis examined in this work is the existence of a correspondence that is both **temporal and directional** between a terrestrial journey and an observable celestial motion.

The account in Matthew indicates that the Magi leave Jerusalem after their meeting with Herod, follow the star they had previously identified, and reach Bethlehem at the moment when it is perceived as "stopping" above the destination.

The aim of this section is to formalize this intuition as an explicit kinematic problem, allowing a quantitative comparison between:

- constraints related to terrestrial motion (distance, duration, orientation), and
- astronomical data describing the position and apparent motion of the Jupiter–Saturn conjunction.

### 7.2. Minimal assumptions and modeling framework

The analysis continues to rely on the minimal working hypotheses introduced in Section 3.2.

### 7.3. Geographical data and terrestrial parameters

The model considers two well-defined locations:

- **Point A**: Jerusalem
- **Point B**: Bethlehem

The distance between these locations is estimated at 8–8.5 km, based on historically plausible routes.

The most likely mode of travel is a terrestrial caravan, with an average speed typically between 4 and 7 $km \cdot h^{-1}$ depending on load, terrain, and conditions (see note 3). This refined range implies a travel time of approximately 1 h 15 min to 1 h 45 min, consistent with the order-of-magnitude estimate given in Section 5.2.

A reference value of 6 $km \cdot h^{-1}$ is adopted as a central estimate (see Note 3).

### 7.4. Definition of the celestial reference frame

To formalize the correspondence between terrestrial motion and celestial motion, a simple reference frame is defined (see Fig. 4):

- **Point E**: apparent position of the conjunction at its first early-evening visibility
- **Point E′**: direction of true south as observed from Jerusalem, corresponding to the Jerusalem–Bethlehem axis

The problem can then be stated as follows:

Do there exist dates for which the time required for the caravan to travel from Jerusalem to Bethlehem matches the time taken by the celestial object to move from its first post-sunset visibility to its alignment with the southern direction?

This formulation defines a **temporal and directional synchronization condition** between terrestrial and celestial motion.

### 7.5. Sky–ground synchronization condition

During the time required for the caravan to travel from point A (Jerusalem) to point B (Bethlehem), the apparent position of the conjunction must evolve from an initial position E (first early-evening visibility or shortly thereafter) to a position E′ aligned with the A–B axis.

This evolution must:

- remain compatible with a continuous perception of directional guidance, and
- culminate in a stationary or quasi-stationary configuration upon arrival

The central condition of the model can be written as:

$$\Delta t(A \rightarrow B) = \Delta t(E \rightarrow E')$$

where:

$\Delta t(A \rightarrow B)$ is the terrestrial travel time

$\Delta t(E \to E')$ is the time required for the celestial object to evolve from E to E′

In this framework, the **only free parameter** is the average speed of the caravan. All other quantities (planetary positions, apparent motion, celestial geometry, and geographical orientation) are independently determined.

The synchronization condition therefore does not rely on arbitrary adjustment, but on strict compatibility between a predetermined celestial dynamics and a constrained terrestrial motion.

## 7.6. Computational method

Initial evaluations were performed manually on a limited set of dates and speeds. While sufficient to validate the conceptual framework, this approach did not allow exhaustive exploration of the solution space.

A dedicated computational tool was therefore developed to systematically evaluate, for each day in the relevant period, the compatibility between terrestrial motion and the apparent motion of the Jupiter–Saturn conjunction.

For each tested date, the program computes:

- the time of first visibility (first early-evening visibility)
- the apparent angular position of the conjunction at regular intervals
- the temporal correspondence between celestial evolution and the A–B displacement

The tool was designed to allow independent reproducibility using the same ephemerides and geographical parameters.

## 7.7. Simulation results

Simulations were performed over the interval from 21 March 7 BCE to 15 January 6 BCE, covering the entire duration of the conjunction.

The parameters used are:

- distance: 8–8.5 km
- average speed: ~6 $km \cdot h^{-1}$ (see Note 3)
- angular tolerance around the south axis: ±1.5° (see Section 8.2)

For each date, the algorithm evaluates:

1. travel time between Jerusalem and Bethlehem
2. evolution of the apparent angular position from first post-sunset visibility to south alignment
3. presence of a planetary station at arrival

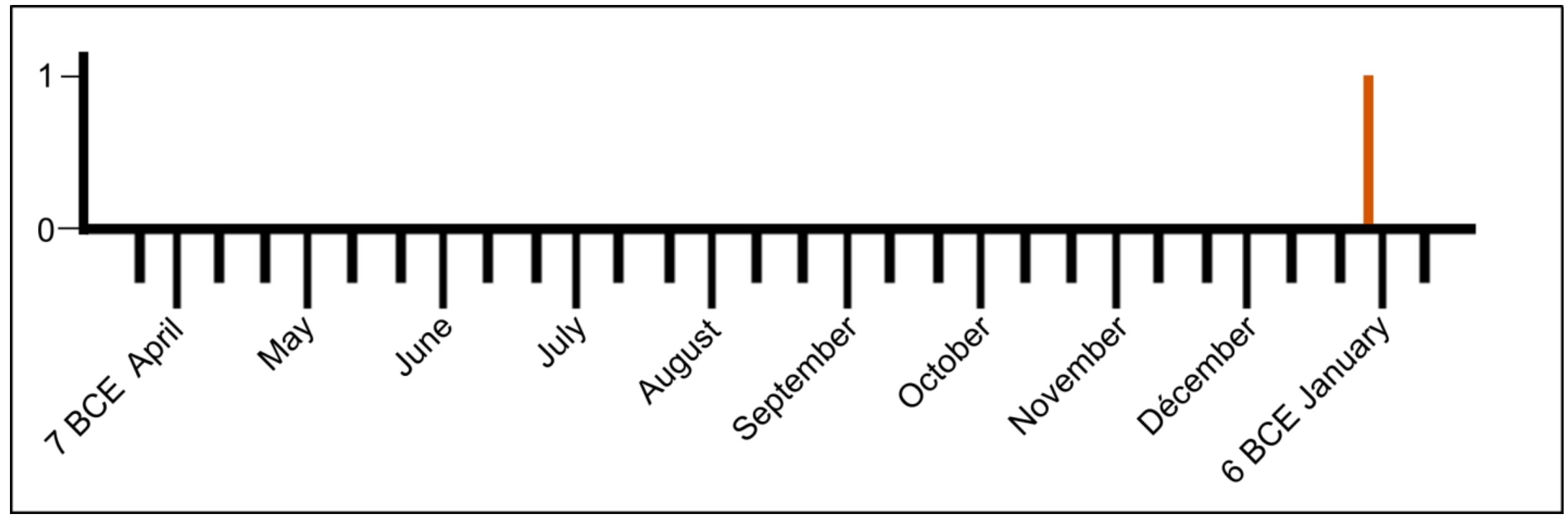

**Figure 5 — Temporal window compatible with kinematic sky–earth synchronization.**
Histogram showing the dates between 21 March 7 BCE and 15 January 6 BCE (Julian calendar) for which the synchronization condition between the terrestrial displacement from Jerusalem to Bethlehem and the apparent kinematics of the Jupiter–Saturn conjunction has been evaluated. Non-zero bars correspond exclusively to the four consecutive dates from 24 to 27 December 7 BCE, which are the only ones satisfying simultaneously all model constraints under the adopted parameters: travel duration consistent with a displacement speed of 6 $km \cdot h^{-1}$, directional alignment with the southward axis, and temporal proximity to a planetary stationary phase. All other dates within the studied interval do not satisfy these conditions for the given parameters and are represented by zero values.

The results show a clear convergence: only four consecutive dates satisfy all constraints simultaneously under the adopted parameters.

These dates are:

- 24 December 7 BCE
- 25 December 7 BCE
- 26 December 7 BCE
- 27 December 7 BCE

All other dates fail to satisfy at least one constraint.

These dates emerge from the intersection of independent constraints (visible after sunset, travel duration, and geometric alignment), rather than from any single isolated astronomical event.

Given that Jupiter is observable only within a limited early-evening window during this period (see Section 4.5), is constrained within a limited observational window rather than freely adjustable.

Figure 5 presents a binary representation of the results over the entire period: only these four dates yield non-zero outcomes.

The robustness of these compatible solutions with respect to variations in travel velocity is illustrated in Figure 6.

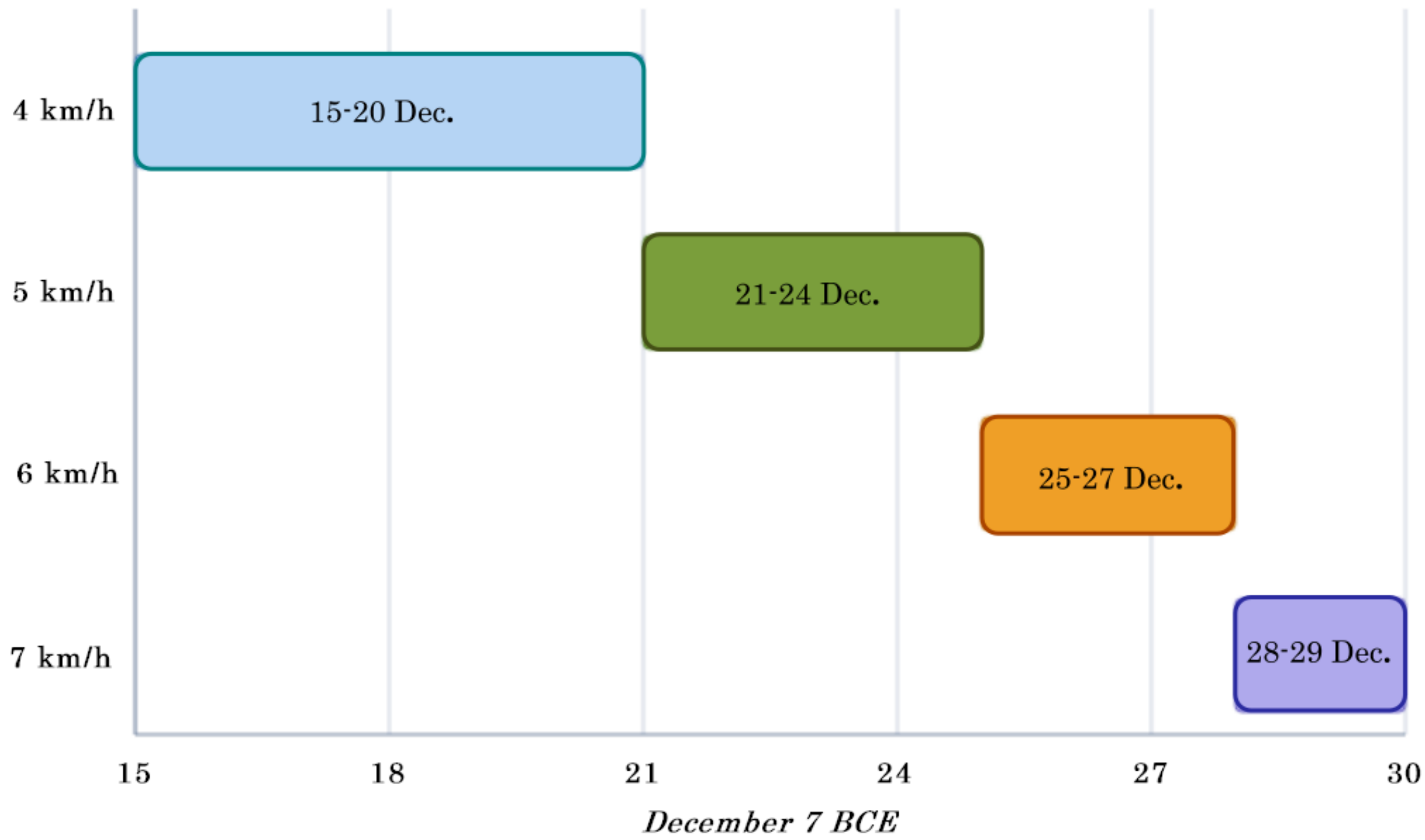

**Figure 6 — Extension of the compatibility domain as a function of travel speed.**
Figure 6 extends the analysis to travel speeds between 4 and 7 km·h$^{-1}$. The set of compatible solutions expands slightly but remains confined to a narrow interval between 15 and 29 December 7 BCE, demonstrating the robustness of the result with respect to variations in the kinematic parameter.

### 7.8. Robustness and structural rigidity of the model

- Variations in the average travel speed within realistic bounds do not significantly alter the structure of the results. The model consistently converges toward a narrow temporal window in late December 7 BCE.
- This concentration results directly from the structure of the model:
    - a single adjustable parameter (travel speed)
    - multiple constraints treated as independent (geometric, kinematic, chronological)
    - absence of ad hoc assumptions

Imposing arbitrary dates leads to incompatibility between terrestrial and celestial constraints, excluding those solutions within the model.

### 7.9. Scope of the result

The result presented here constitutes the analytical core of the study.

It does not aim, by itself, to establish a historical fact, but to demonstrate the existence of a **precise and non-trivial correspondence** between the narrative of Matthew and a well-defined astronomical phenomenon.

The consistency between textual constraints, terrestrial geometry, and celestial kinematics is not readily attributable to chance alone. It suggests a possible link between the narrative description and a real astronomical event interpreted as meaningful by ancient observers.

A formal probabilistic evaluation of this correspondence is developed in section 8.

### 7.10. Selective perception of the phenomenon

A rarely addressed issue concerns the apparently selective perception of the phenomenon: only the Magi recognize its significance, while others do not.

Many explanations in the literature rely on miraculous or symbolic interpretations, which are not testable.

The present model offers a simpler explanation based on historical and cognitive considerations.

A planetary conjunction is observable by any attentive observer, but its significance depends on the interpretative framework applied to it.

The Magi, described as eastern astrologers, belonged to a category of specialized observers capable of interpreting such configurations. In the cultural context of antiquity, conjunctions involving Jupiter and Saturn could carry political or royal significance.

For non-specialists, the same phenomenon would remain an ordinary celestial event without particular meaning.

Thus, what is selective is not the observation itself, but its interpretation. This distinction accounts naturally for the narrative without introducing discontinuity between the astronomical phenomenon and its human context.

Any alternative hypothesis must therefore explain, in a coherent and non-ad hoc manner, why the phenomenon was interpreted as significant by a restricted group while remaining otherwise unremarkable.

## 8. Constraint-Based Analysis and Assessment of the Role of Chance

**Preliminary Methodological Note**

The approach developed in this section is not a probabilistic[5] analysis in the strict frequentist sense. It does not attempt to estimate the frequency of occurrence of an event within a hypothetical ensemble of repeated trials, as such a framework is not applicable to a unique historical configuration.

Instead, the method consists in a progressive restriction of a well-defined space of admissible configurations through the cumulative application of independent or weakly dependent constraints. Each constraint—temporal, directional, or astronomical—reduces the volume of this space. The resulting quantity is therefore more appropriately interpreted as a *measure of the remaining compatible domain* rather than as a probability in the strict sense.

In this framework, the product of the retained fractions provides an **order-of-magnitude estimate of the degree of constraint** imposed by the simultaneous satisfaction of all conditions. This approach is analogous to constraint-satisfaction analyses commonly used in inverse problems, where the objective is to evaluate how strongly a solution is restricted by independent criteria.

[5] The probabilistic estimates presented in this work do not correspond to strict frequentist probabilities. They are intended as order-of-magnitude indicators of the reduction of the admissible configuration space under multiple constraints, rather than as exact measures of event likelihood.

The numerical values derived below should therefore be interpreted as **compatibility fractions** or **compatibility indicators**, and not as normalized probabilities defined over a rigorously constructed sample space.

## 8.1. Rationale for a Constraint-Based Quantification

The existence of a precise correspondence between an astronomical configuration and a historical narrative naturally raises the question of whether such a match could arise by chance.

In the absence of a well-defined stochastic process generating comparable events, the relevant question is not the frequency of occurrence of the observed configuration, but rather:

*To what extent does the simultaneous satisfaction of all constraints restrict the space of admissible configurations?*

The objective of this section is therefore to quantify, at the level of orders of magnitude, how strongly the identified constraints reduce the set of possible solutions.

## 8.2. Definition of the Configuration Space

A "configuration" is defined as a set of astronomical and terrestrial parameters satisfying, at least partially, the constraints derived from the narrative. These parameters include:

- the date of occurrence;
- the duration of visibility;
- the apparent celestial position;
- the existence of a directional motion;
- the presence of a stationary phase;
- chronological compatibility with the reign of Herod;
- geographical compatibility with a Jerusalem–Bethlehem route.

The relevant configuration space is therefore not the full set of astronomical phenomena, but a restricted subset satisfying minimal physical, observational, and historical conditions.

This construction explicitly excludes:

- transient or instantaneous phenomena;
- unpredictable or non-repeatable events;
- configurations not observable from the latitude of Judea;
- phenomena lacking both directional coherence and apparent stationarity.

The resulting space is thus already highly structured prior to any quantitative assessment.

In this framework, the set of admissible configurations can be viewed as a reduced region within a higher-dimensional parameter space ("phase space"), whose dimensions correspond to the relevant astronomical, temporal, and geometrical variables. Each constraint acts as a projection or restriction on this space, progressively reducing its effective volume. The final solution identified in this study thus corresponds to a small connected subset of this parameter space, whose relative measure provides a natural quantitative indicator of its degree of constraint.

## 8.3. Order-of-Magnitude Constraint Analysis

The constraint-based estimate is obtained by considering three largely independent filtering criteria.

### 8.3.1. Temporal Constraint Fraction

For velocities in the interval 4–7 $km \cdot h^{-1}$, the model yields 15 compatible dates within a total interval of 301 days.

$$\frac{\mathbf{15}}{\mathbf{301}} \approx \mathbf{5{,}0 \times 10^{-2}}$$

This fraction represents the proportion of the explored temporal domain satisfying the kinematic synchronization constraint.

### 8.3.2. Directional Constraint Fraction

The admissible angular tolerance (±1.5°)[6] corresponds to a total window of 3° over a 360° horizon.

$$\frac{\mathbf{3}}{\mathbf{360}} \approx \mathbf{8{,}3 \times 10^{-3}}$$

### 8.3.3. Astronomical Rarity Constraint

A triple Jupiter–Saturn conjunction occurs approximately every 855 years. Restricting to a 5-year admissible interval yields:

$$\frac{\mathbf{5}}{\mathbf{855}} \approx \mathbf{5{,}8 \times 10^{-3}}$$

### 8.3.4. Combined Constraint Measure

Assuming approximate independence[7] at first order, the combined reduction factor is:

$$\boldsymbol{P_{global}} \approx \mathbf{2 \times 10^{-6}}$$

This quantity should be interpreted as an **order-of-magnitude measure of the residual compatible domain** after application of all constraints.

## 8.4. Interpretation

The resulting constraint measure indicates that only a very small fraction of the admissible configuration space simultaneously satisfies all conditions.

While this does not constitute a probabilistic proof in the strict sense, it demonstrates that the observed concordance is **highly non-generic** within the defined space. In other words, the solution occupies a narrowly constrained region rather than arising as a typical or expected outcome.

6 The angular tolerance of ±1.5° corresponds to a practical estimate of naked-eye directional resolution near the horizon. This value is consistent with ranges commonly adopted in archaeoastronomical studies. Sensitivity tests performed within ±1°–±2° yield identical compatible dates, indicating that the result is not dependent on the precise choice of this parameter.

7 The assumption of independence between the different constraints is an approximation adopted for the purpose of order-of-magnitude estimation. While partial correlations may exist, they do not affect the qualitative conclusion regarding the highly constrained nature of the solution.

### 8.5. Additional Restrictive Factors

Several additional constraints further reduce the effective solution space:

- limited daily visibility windows;
- observational uncertainty bounds;
- restricted velocity range compatible with historical travel conditions.

These factors do not fundamentally alter the order of magnitude of the estimate but reinforce the conclusion that the compatible domain is extremely small.

### 8.6. Limitations

This analysis relies on several simplifying assumptions:

- independence between constraints is only approximate;
- the configuration space is model-dependent;
- cultural and symbolic interpretations are not included.

However, sensitivity tests indicate that relaxing these assumptions within reasonable bounds does not significantly modify the resulting order of magnitude.

### 8.7. Methodological Position

The constraint-based analysis is not intended to replace historical or astronomical arguments. Its role is to provide a quantitative indicator of how strongly the solution is constrained.

It therefore serves as a complementary argument, highlighting the non-trivial character of the correspondence and motivating further examination of alternative hypotheses.

## 9. Reproducibility, Data, and Independent Verification

### 9.1. Principle of Reproducibility

A fundamental criterion of any scientific approach is the ability for independent verification by third parties. The model proposed here explicitly satisfies this requirement: it relies on no hidden parameters, no proprietary data, and no arguments from authority.

The entire reasoning is based on accessible astronomical data, elementary geometric relationships, and simple kinematic assumptions, all of which can be evaluated independently of the interpretative framework adopted.

### 9.2. Astronomical Data Used

The astronomical data employed in this study are exclusively derived from accessible sources and explicitly described processing steps. They include:

- back-calculated ephemeris tables available in the specialized literature and standard astronomical databases, used to determine apparent positions, retrograde phases, and station dates of the relevant planets;

- a synthetic ephemeris table constructed by the authors, covering the period from March 7 BCE to January 6 BCE, incorporating the parameters required for analyzing visibility, angular separation, and kinematic synchronization between terrestrial motion and the apparent motion of the Jupiter–Saturn conjunction.
  This table, provided as a spreadsheet in the associated Zenodo repository (Bodor & Bauduin, 2026; https://doi.org/10.5281/zenodo.19153655), is directly derived from published astronomical data and contains no a posteriori adjusted parameters. Its purpose is to facilitate independent verification of the results and reproduction of the figures presented in previous sections.

All data used in this study are publicly available.

## 9.3. Independent Verification

The results presented in this study can be independently verified by any reader with access to standard astronomical data and common computational tools. The approach requires no proprietary software and no undocumented implicit assumptions.

Reproducing the analysis involves:

- recalculating planetary positions from published ephemerides;
- reconstructing the Jerusalem–Bethlehem route and its geographic orientation;
- evaluating visibility conditions and kinematic synchronization according to the explicitly defined criteria of the model.

The observed agreement arises from objective and measurable constraints. Any discrepancies can therefore be discussed transparently by identifying the specific assumption or parameter involved.

Importantly, reproducibility does not depend on acceptance of the interpretative hypothesis. A skeptical reader applying the same data and constraints should obtain the same numerical results, regardless of their historical or symbolic interpretation.

## 9.4. Key Validation Condition

The central condition of the model is the following:

the time required to travel the distance from Jerusalem to Bethlehem must be comparable to the time taken by the celestial object to move from its first nocturnal visibility to its transit across the southern meridian.

For an average travel speed of 6 $km \cdot h^{-1}$, the travel time is approximately 85 minutes for a distance of about 8.5 km.

This value follows from a simple kinematic calculation:

$$\frac{(\mathbf{8,5}\ \boldsymbol{km})}{(\mathbf{6}\ \boldsymbol{km} \cdot \boldsymbol{h}^{-\mathbf{1}})} \cong \mathbf{1,42}\boldsymbol{h} \cong \mathbf{85}\ \boldsymbol{min}.$$

This duration can be recomputed for other reasonable speeds without substantially altering the conclusion. The validation condition therefore consists in identifying the dates for which the time interval between first visibility of the conjunction and its transit across the southern meridian is of this order of magnitude.

## 9.5. Result of Independent Verification

Analysis of the kinematic parameters associated with the selected configuration allows identification of the corresponding period. In the absence of any angular tolerance range, inversion of the model yields a highly constrained solution centered on the night of December 24–25, 7 BCE.

This configuration results from the close agreement between the apparent position of the Jupiter–Saturn conjunction, its stationary or near-stationary state, and the synchronization required with the terrestrial motion considered. It is therefore not obtained by selecting from a broad time interval, but by directly identifying the period during which all model parameters are simultaneously satisfied.

Notably, this window coincides with the immediate post-stationary phase of Jupiter identified in Section 4.6. At the moment when kinematic synchronization is achieved, Jupiter has just completed its reversal of motion and exhibits an almost zero apparent angular velocity—consistent with the Matthean description of the star "stopping."

This convergence between two independent analyses, based on the same ephemerides but different approaches, reinforces the internal coherence of the model.

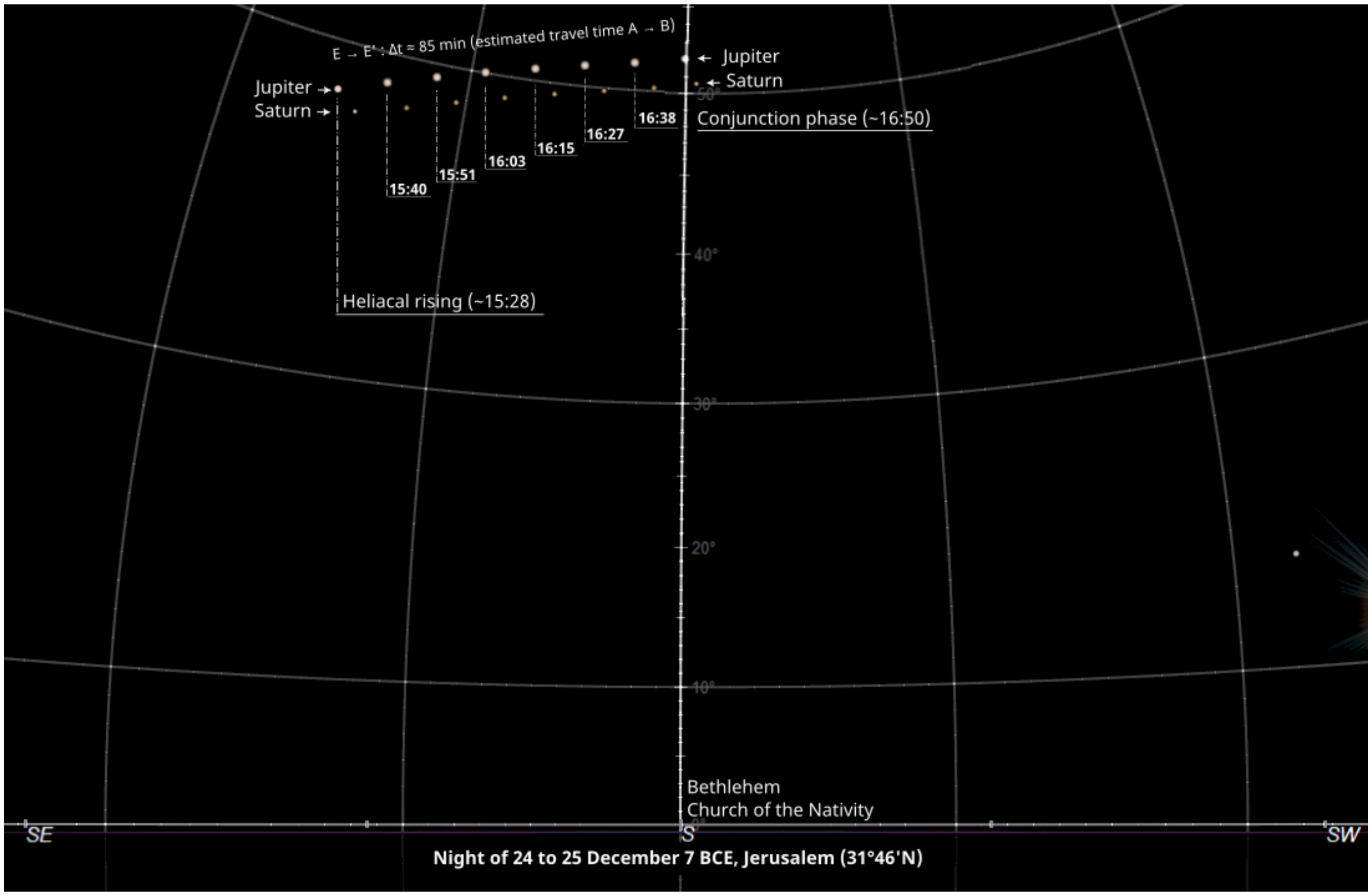

**Figure 7 — Reconstruction of the sky from Jerusalem (31°46′N) on the night of 24–25 December 7 BCE (Julian calendar), generated using the Starry Night Pro Plus 8 software.**

The Jupiter–Saturn conjunction is shown at the time of its first post-sunset visibility (E, ~15:28, southeastern sector), as well as during its apparent progression until the conjunction phase (~16:50, direction of the southern meridian). Successive positions are sampled at equal time intervals to illustrate the apparent continuity of motion and do not correspond to independent observational measurements. The interval $\Delta t \approx 85$ min is consistent with the estimated travel time of a caravan between Jerusalem (A) and Bethlehem (B). In this configuration, the direction of Bethlehem (Church of the Nativity) is aligned with the southern meridian at the time corresponding to arrival, illustrating a geometric and temporal configuration compatible with a directional interpretation of the phenomenon during terrestrial movement.

Comparable sky configurations and visibility timings were independently reproduced in Starry Night Pro 5, Starry Night Pro 8, and Stellarium after normalization of daylight-saving and local time settings, reinforcing the operational robustness of the adopted visibility criterion.

## 9.6. Robustness of the Result

Sensitivity tests were conducted by varying several model parameters, including:

- average travel speed within a realistic range,
- observer latitude,
- astronomical visibility criteria.

In all cases, the solution remains confined to a narrow temporal window centered on late December 7 BCE. No other period was identified with a comparable level of agreement across all geometric, kinematic, and astronomical constraints.

## 9.7. Temporal Consistency of the Travel Scenario

Beyond identifying a compatible calendar window, the ephemerides allow examination of the internal temporal coherence of the scenario corresponding to the central date (December 24–25, 7 BCE, at 6 $km \cdot h^{-1}$).

On that date, the conjunction becomes visible from Jerusalem at dusk (~15:28)[8], at an altitude of approximately 47° and an azimuth toward the southeast. This timing is consistent with a late-afternoon departure following the meeting with Herod, given the short winter daylight hours at this latitude.

Assuming departure shortly after first visibility and a travel speed of 6 $km \cdot h^{-1}$, arrival in Bethlehem occurs around 16:50, corresponding to the conjunction's transit across the southern meridian.

At that time, night has already fallen, yet the arrival remains compatible with an early evening visit, without requiring implausible or socially atypical timing.

Importantly, this temporal coherence was not imposed a priori. A mismatch between terrestrial travel time and celestial motion could easily have produced an implausibly late arrival. The fact that synchronization naturally yields a coherent astronomical, geometrical, and human timeline constitutes an additional consistency of the model, without introducing new assumptions.

## 9.8. Local Topographical Constraints

The model is based on an idealized representation of the journey and does not explicitly incorporate local topographical and architectural constraints[9].

---

[8] In historical astronomical convention, the Julian day begins at noon rather than at midnight. All times reported in this work follow the modern civil convention (midnight as the start of the day). Accordingly, the date "25 December 7 BCE" refers to the civil day beginning at midnight, consistent with the Julian calendar in use at the time.

[9] The estimated distance required to clear the built environment surrounding Herod's palace (~400–500 m) is based on archaeological reconstructions of the site. This adjustment slightly reduces the effective travel distance after first visibility, without affecting the overall conclusions of the model.

At departure from Herod's palace, the Magi would have been in a dense built environment with high walls, limiting visibility of the sky. Since the Gospel suggests that the star is observed at that moment (Mt 2:10), the observation must reasonably occur only after reaching a sufficiently open area.

Subtracting the minimal distance required to clear the Herodian complex (~445 m), the effective post-observation distance is reduced to approximately 7.5–8 km, yielding a revised average speed of 5.2–5.6 $km \cdot h^{-1}$.

This corrected value lies well within the robustness range established in Section 7.8 (4–7 $km \cdot h^{-1}$). The topographical correction therefore does not alter the identified temporal window; it confirms it with a slightly lower but fully compatible speed.

For this reason, it is treated as a secondary refinement rather than a modification of the core model.

### Conclusion of Section 9

Taken together, these elements demonstrate that the analysis is fully reproducible from public data using standard tools and explicitly stated assumptions.

The following section examines the implications, limitations, and possible interpretations of these results.

## 10. Discussion

### 10.1. Scope of the Result

No date is fixed *a priori* in the model: all chronological results emerge exclusively from the confrontation between independent astronomical data and explicitly defined kinematic constraints.

The model does not rely on any assumption regarding the theological intent of the text or its intrinsic historical value; it tests only its descriptive compatibility with a real astronomical phenomenon, without postulating causality or intentional meaning.

The result presented here does not constitute a historical proof nor an exhaustive explanation of the narrative, but rather the identification of a non-trivial kinematic and temporal correspondence, sufficient to constrain a broad class of competing hypotheses.

The main result of this work highlights a precise temporal and directional correspondence between the terrestrial motion described in Mt 2:1–12 and the apparent kinematics of the great Jupiter–Saturn conjunction of 7 BCE. This correspondence emerges only within an extremely narrow temporal window, centered on late December of that year.

It is important to emphasize that this agreement does not result from *a posteriori* adjustment, but from the simultaneous satisfaction of independent constraints:

- geographic (Jerusalem–Bethlehem orientation),
- kinematic (realistic travel speed),
- astronomical (visibility and transit times),
- temporal (overall duration of the conjunction).

The emergence of a unique solution within such a constrained parameter space constitutes, in itself, a non-trivial methodological result.

### 10.2. Relation to Kepler's Intuition

As early as the 17th century, Johannes Kepler proposed that the "Star of Bethlehem" could be related to a major planetary conjunction, in particular the Jupiter–Saturn conjunction of 7 BCE. His hypothesis was primarily based on astrological and symbolic considerations, given the absence of computational tools allowing detailed quantitative verification.

The present work does not adopt Kepler's astrological framework. Instead, it reformulates the problem in strictly kinematic and geometric terms. In this sense, it does not validate Kepler at the level of symbolic interpretation, but it does support his fundamental intuition that this particular conjunction deserves specific attention in any attempt to understand the Matthean account.

### 10.3. Compatibility with the Text of Matthew

A central issue concerns the relationship between the model and the Gospel text. It is important to stress that the model does not aim to "explain" Matthew, nor to correct or modernize it. It tests a minimal hypothesis: is the narrative compatible with a real astronomical phenomenon, without violating the known laws of celestial mechanics?

The problematic expressions—"the star went before them" and "it stopped over the place"—find, within the proposed model, a coherent descriptive interpretation, without requiring *ad hoc* supernatural assumptions.

In this perspective, Matthew's language appears as that of a narrative reporting a perceived phenomenon, rather than that of an astronomer seeking technical precision.

### 10.4. Limitations of the Model

Despite its internal consistency, the proposed model has limitations that must be explicitly acknowledged.

First, it relies on a methodological assumption: that the Gospel of Matthew refers to a real event rather than a purely symbolic construction. This assumption is neither demonstrated nor refuted here; it is simply tested for consistency.

Second, certain logistical assumptions (travel speed, continuity of motion) are necessarily simplified. While the orders of magnitude are realistic, they do not claim to capture the full complexity of an ancient journey.

Third, the model does not attempt to identify the precise geographical origin of the Magi. It only shows that even a long-distance journey remains compatible with the unusually long duration of the conjunction.

These limitations do not undermine the internal coherence of the model, but they explicitly delimit its domain of validity.

### 10.5. Possible Objections and Methodological Response

A recurrent objection concerns the *a posteriori* character of the analysis: given the traditional date of Christmas, the model might be suspected of artificially converging toward that date.

This objection does not withstand methodological scrutiny. The model contains no temporal parameter fixed *a priori*. The dates emerge exclusively from the interaction between astronomical data and kinematic constraints. The fact that the solution lies near a later tradition is therefore a result, not an initial assumption.

Another objection concerns the independence of the probabilistic factors combined in the analysis. It is true that any probabilistic reasoning applied to a historical event must be handled with caution. The probabilities presented here are not intended as absolute measures, but as order-of-magnitude indicators of the degree of constraint imposed by the model.

### 10.6. Interdisciplinary Implications

This work lies at the intersection of several disciplines:

- astronomy and celestial mechanics,
- ancient history,
- biblical exegesis,
- philosophy of science.

It illustrates the value of an interdisciplinary approach when methodological boundaries are clearly respected. Planetary science does not "prove" history, but it can strongly constrain the range of plausible interpretations of an ancient text.

### 10.7. Summary of the Discussion

The discussion highlights an essential point: the Gospel of Matthew, often considered problematic from a scientific standpoint, is shown to remain consistent under rigorous quantitative analysis, provided it is approached without prior bias, either favorable or unfavorable.

The result obtained does not settle the debate on the Star of Bethlehem, but it reframes it on a more precise ground, where arguments can be evaluated, reproduced, and discussed in a rational manner.

## 11. General Conclusion and Perspectives

### 11.1. General Conclusion

The aim of this work was to examine, using tools from modern astronomy and kinematic modeling, the hypothesis that the "Star of Bethlehem" described in Mt 2:1–12 could be associated with the great Jupiter–Saturn conjunction of 7 BCE—an idea first proposed in the early 17th century by Johannes Kepler.

The analysis reveals the existence of a coherent correspondence, within the set of considered constraints, between:

- the apparent kinematics of this conjunction,
- the geometry of the Jerusalem–Bethlehem route,
- the temporal and logistical constraints of realistic human travel,
- and certain descriptive elements of the Matthean narrative.

This correspondence emerges within a restricted temporal window, centered on late December 7 BCE. Within the framework of the adopted assumptions, the probabilistic assessment suggests that such an agreement is non-trivial.

It is important to emphasize that this analysis does not aim to establish a historical fact in the strict sense, nor to validate a religious interpretation. Rather, it shows that a real, rare, and datable astronomical configuration can be considered compatible with a reading of the narrative in a geometric and kinematic framework, without invoking additional assumptions.

### 11.2. Main Contributions

The specific contributions of this study may be summarized as follows:

1. **Falsifiable methodology**
   The Gospel narrative is treated as a set of testable constraints, rather than as a text to be interpreted symbolically or defended *a priori*.
2. **Original kinematic modeling**
   The motion of the Magi is explicitly related to the apparent motion of the conjunction, allowing a direct confrontation between terrestrial and celestial dynamics.
3. **Quantitative re-examination of Kepler's intuition**
   The Keplerian hypothesis is analyzed using modern computational tools, enabling a fine-grained quantitative verification at the terrestrial level.
4. **Probabilistic framing of the result**
   The probabilistic analysis provides an order-of-magnitude estimate of the model's degree of constraint, reinforcing the robustness of the conclusion.

### 11.3. Limitations and Interpretative Caution

As with any interdisciplinary study, this work has limitations that must be acknowledged:

- It relies on the minimal assumption that the text of Matthew refers to an event perceived as real by its author or source, without prejudging its nature.
- The logistical parameters employed represent realistic averages, not an exhaustive reconstruction of historical conditions.
- The probabilities presented are not absolute measures, but indicators of rarity and constraint.

These limitations do not undermine the internal coherence of the model, but they define its domain of validity.

### 11.4. Perspectives

Several natural extensions of this work can be considered:

- a comparative analysis with other historical great conjunctions, in order to better characterize the uniqueness of the 7 BCE event;
- a detailed linguistic study of the Greek and Syriac versions of Mt 2:1–12, in relation to descriptive astronomical terminology;
- a methodological extension to other ancient narratives involving celestial phenomena, within a framework of scientific source criticism.

More broadly, this work invites a reconsideration of the often-assumed boundary between ancient texts and the exact sciences. When a text is formulated with sufficient precision, it may, in some cases, be subjected to rigorous quantitative analysis.

### 11.5. Final Remark

The question raised by Johannes Kepler more than four centuries ago—the nature of the Star of Bethlehem—is neither anecdotal nor definitively resolved. It represents a characteristic problem at the interface between history, science, and rational inquiry.

Without taking a position on theological grounds, this work shows that a rigorous dialogue between an ancient text and the tools of modern science is not only possible, but can prove intellectually fruitful.

### 11.6 Acknowledgments

The authors gratefully acknowledge Heino Falcke for helpful comments and suggestions on an earlier version of this work.